\documentstyle[prd,aps,preprint,epsfig]{revtex}
\newfont\fiverm{cmr5}
\input prepictex
\input pictex
\input postpictex

\begin{document}

\newcommand{\TeV}{\,{\rm TeV}}
\newcommand{\GeV}{\,{\rm GeV}}
\newcommand{\MeV}{\,{\rm MeV}}
\newcommand{\keV}{\,{\rm keV}}
\newcommand{\eV}{\,{\rm eV}}
\def\ap{\approx}
\def\bea{\begin{eqnarray}}
\def\eea{\end{eqnarray}}
\def\beqar{\begin{eqnarray}}
\def\eeqar{\end{eqnarray}}
\def\ler{\lesssim}
\def\gtr{\gtrsim}
\def\beq{\begin{equation}}
\def\eeq{\end{equation}}
\def\haf{\frac{1}{2}}
\def\plb#1#2#3#4{#1, Phys. Lett. B {\bf #2} (#4) #3}
\def\npb#1#2#3#4{#1, Nucl. Phys. {\bf B#2} (#4) #3}
\def\prd#1#2#3#4{#1, Phys. Rev. D {\bf #2} (#4) #3}
\def\prl#1#2#3#4{#1, Phys. Rev. Lett. {\bf #2} (#4) #3}
\def\mpl#1#2#3#4{#1, Mod. Phys. Lett. A {\bf #2} (#4) #3}
\def\rep#1#2#3#4{#1, Phys. Rep. {\bf #2} (#4) #3}
\def\lpp{\lambda''}
\def\ccg{\cal G}
\def\slash#1{#1\!\!\!\!\!/}
\def\rpv{\slash{R_p}}

\setcounter{page}{1}
\draft
\preprint{hep-ph/0103064}

\title{Implications of the muon anomalous magnetic moment and
Higgs-mediated flavor changing neutral currents }

\author{Sin Kyu Kang and  Kang Young Lee }

\address{School of Physics,
Korea Institute for Advanced Study, Seoul 130-012, Korea}


\tighten

\maketitle

\begin{abstract}
In the light of the recent measurement of the muon 
anmalous magnetic moment $a_\mu$ by the Muon
$(g-2)$ Collaboration,
we examine the contribution to $a_\mu$
from the exchange of flavor changing scalars.
Assuming that the heavier generations have larger flavor
changing couplings, we obtain a bound on $\mu - \tau$ Yukawa
coupling for a given scalar mass.
Constraints on other flavor changing/conserving couplings are 
also obtained from the lepton flavor violating decays of 
muon and tau lepton,
and bounds on the branching ratio of $\tau \rightarrow 3 e$,
$\mu \rightarrow 3 e$ and $\tau \rightarrow e \mu e$ processes
are predicted.

\end{abstract}



Very recently, the Muon $(g-2)$ Collaboration has reported
a measurement of the anomalous magnetic moment of the muon,
indicating a $2.6\sigma$ deviation of $a_{\mu}\equiv
(g_{\mu}-2)/2$ from the standard model (SM) prediction \cite{muon}:
\beq
\label{deviation}
\Delta a_{\mu}\equiv a_{\mu}^{\rm exp}-a_{\mu}^{\rm SM}=
(426\pm 165)\times 10^{-11}.
\eeq
Although the measurement of $a_{\mu}$ is about 350 times less precise
than that of $a_{e}$, it is much more sensitive to new physics effects
since such contributions are generally proportional to $m^2_{l}$ 
\cite{amu,marciano}.
In this spirit, the deviation exhibited in Eq. (\ref{deviation})
may be regarded as a signal of new physics beyond the SM,
and at $90\%$ C.L., $\Delta a_{\mu}^{NP}$ must lie in the range
\beq
\label{dev90}
 215 \times 10^{-11} \lesssim 
            \Delta a_{\mu}^{NP} \lesssim 637 \times 10^{-11}.
\eeq
Various proposals of new physics have been suggested
to accommodate this deviation, and it has also been employed to constrain
the unknown parameters \cite{proposal}.

In this paper, we will examine another possibility that stems from
Higgs-mediated lepton flavor violating neutral interactions 
which contribute to the anomalous magnetic moment of the muon.
In general the flavor changing neutral current (FCNC) interactions 
can be generated at tree level in the two-Higgs doublet extension of the SM.
Since such tree level flavor changing neutral current
interactions are phenomenologically dangerous, 
one usually impose a discrete symmetry  
in order to avoid the FCNC problem. 
However, no tree level FCNC may not be necessary,
alternatively one can consider the very general model
with extended Higgs sector 
where the flavor changing neutral couplings are just constrained 
by experiments.
Nie and Sher have studied the effects of 
those Higgs-mediated FCNC couplings on $a_{\mu}$, 
and shown that contribution from the exchange
of flavor-changing scalars can be enhanced 
as long as the mass of scalar is light enough \cite{sher1}.  
Reliably assuming that the heavier generations have 
larger flavor changing couplings \cite{sher2}, 
they have obtained a bound on $\mu - \tau$ Yukawa coupling 
from the measured value of $a_{\mu}$.
In this work, we will re-analyze the effects of the lepton flavor
changing (particularly $\mu - \tau$) coupling on $a_{\mu}$
to obtain an improved bound on the coupling
in the light of new measurement of $a_\mu$. 
On the other hand, such FCNC couplings generically lead to 
the lepton flavor violating decays of $\tau $ and $\mu $ 
which should be strongly constrained by experimental data.
Combining the constraint on the coupling obtained from Eq.(\ref{dev90})
with experimental limits on the branching ratios of the
corresponding processes,
it is possible to estimate bounds on other flavor changing/conserving 
couplings, $e-e$, $e-\mu$, $e-\tau$, $\mu-\mu$, and $\tau-\tau$.
We will investigate the bounds on those couplings and the branching
ratios for lepton flavor violating three-body decays of $\tau$.

Following Ref. \cite{sher1}, we choose a basis for the
two Higgs doublets, $H$ and $\Phi$, 
such that only one Higgs doublet, $H$, 
obtains a vacuum expectation value. 
After the rotation of the charged leptons,
corresponding neutral Higgs boson $H^0$ has flavor-diagonal couplings, 
while the other neutral Higgs boson, $\phi$, 
have flavor changing couplings.
The relevant terms for Yukawa interaction in the Lagrangian are then given 
by \cite{sher1}
\beq
\label{lagrangian}
L = \frac{\sqrt{2}}{v}
    \left( m_e \overline{e}_{L}e_R H^0 
         + m_{\mu} \overline{\mu}_{L} \mu_R H^0
         + m_{\tau} \overline{\tau}_{L}\tau_R H^0 \right) 
    + h_{ij} \overline{l}_{iL}
     l_{jR} \phi + H.c.
\eeq
where $\langle H \rangle = (0, v/\sqrt{2})^T$, 
and the $\phi$ field consists of a scalar $\phi_s$ and a pseudoscalar
$\phi_p$. 
The lepton flavor changing couplings can induce new contributions to
the anomalous magnetic moment of the muon, $a_{\mu}$, as well as the lepton
flavor violating decays of the muon and tau
through scalar and pseudoscalar-mediated diagrams.
As shown in Ref.\cite{sher2}, the heavier generations
are naturally expected to have larger flavor-changing couplings.
Then, the most dominant contributions to the anomalous magnetic moment of the
muon arise from $h_{\mu \tau}$ coupling.
The scalar and pseudoscalar contributions to $a_\mu$
are given by \cite{sher1}
\beq
\label{muon1}
a^{s,p}_{\mu} = \frac{h^2_{\mu \tau}m^2_{\mu}}{16\pi^2}
        \int^{1}_{0} dx \frac{x^2-x^3\pm (m_{\tau}/m_{\mu})x^2 }
        {m^2_{\mu} x^2 + (m^2_{\tau} - m^2_{\mu}) x + m^2_{\phi_{s,p}} (1-x)},
\eeq
where $m_{\phi_{s,p}}$ is the mass of the scalar or pseudoscalar, 
and the $+(-)$ sign corresponds to the scalar (pseudoscalar).
In the limit $m_{\mu} \ll m_{\tau} \ll m_{\phi}$, the leading term becomes
\beq
\label{muon2}
a_{\mu} \simeq (\pm) \frac{h^2_{\mu \tau}}{16\pi^2}
\frac{m_{\mu}m_{\tau}}{m_{\phi}^2} \left(\ln \frac{m_{\phi}^2}{m_{\tau}^2}
-\frac{3}{2}\right).
\eeq
We note that the scalar exchange yields a positive contribution to $a_{\mu}$, 
while the pseudoscalar exchange leads to a negative one.
Since our goal is to explain the recent measured value of $a_{\mu}$
which indicates a positive value of $\Delta a_\mu$, 
we demand that the contribution from the scalar exchange dominates 
over that from the pseudoscalar exchange.
This can be easily achieved by assuming that the pseudoscalar 
is sufficiently heavier than the scalar.
We see from Eq.(\ref{muon2}) that $a_{\mu}$ decreases 
as the mass of the scalar increases for a fixed $h_{\mu \tau}$.
For a given $m_{\phi}$, $\Delta a_{\mu}$ yields a constraint on
the coupling $h_{\mu \tau}$.
In particular, when we combine the lower limit of neutral Higgs boson mass
determined from CERN $e^{+} e^{-}$ collider LEP experiment,
the lower value of $\Delta a_{\mu}$ in Eq. (\ref{deviation}) 
leads to the lower bound on $h_{\mu \tau}$.

In order to calculate the contribution to $a_{\mu}$ from the scalar exchange,
let us take the mass of the scalar to be $100$ GeV 
which is about the value of the current LEP bounds, 
and assume that the mass of the pseudoscalar is large enough 
so that its contribution becomes negligibly small.
The value of $a_{\mu}$ then becomes 
$4.65 \times 10^{-8}h_{\mu \tau}^2 $.
Picking up the limit for the measured value of $a_{\mu}$, 
we can get a constraint on $h_{\mu \tau}$ which is given by
$ 0.053 \leq h_{\tau \mu} \leq 0.09$
for $m_{\phi}=100$ GeV.

Now, let us examine the lepton flavor violating processes such as
$\tau \rightarrow \mu \gamma, ~~\tau \rightarrow e \gamma$,
$\mu \rightarrow e \gamma$, and
three-body decays of $\tau$ and $\mu$ which can be induced from
the FCNC couplings.
A matrix element of the form,
\beq
i{\cal M} = e \overline{u}(p_2)~ i \sigma_{\mu \nu}q^{\nu} 
                (A_L P_L + A_R P_R) u(p_1)\epsilon(q)^{\ast \mu},
\eeq
leads to the partial widths for the radiative decay modes which are given
in the leading order by \cite{sher3}
\bea
\Gamma(\tau \rightarrow e \gamma) &\simeq &\frac{\alpha m_{\tau}^5}
{2^{11}9 \pi^4 m_{\phi}^4}(h^2_{\tau \tau} h^2_{e \tau}
+4 h^2_{ee} h^2_{e\tau} + 4 h^2_{\mu \tau} h^2_{e \mu}), \\
\Gamma(\tau \rightarrow \mu \gamma) &\simeq &\frac{\alpha m_{\tau}^5}
{2^{11}9 \pi^4 m_{\phi}^4}(h^2_{\tau \tau} h^2_{\mu \tau}
+4 h^2_{e\mu} h^2_{e \tau} + 4 h^2_{\mu \tau} h^2_{\mu \mu}), \\
\Gamma(\mu \rightarrow e \gamma) &\simeq &\frac{\alpha m_{\mu}^3 m_\tau^2}
{2^{13}\pi^4 m_{\phi}^4}(h^2_{\mu \tau} h^2_{e \tau}) ,
\eea
where we have considered only tau-mediated loop contribution 
for the $\mu \rightarrow e \gamma$ process.
The lepton flavor violating three-body decay rate of 
$\tau $ is given by 
\beq
\Gamma(\tau \rightarrow l_i l_j l_k) \simeq  \frac{m_{\tau}^5}
{3072\pi^3 m_{\phi}^4} (h_{l_j l_k}^2 h_{l_i \tau}^2),
\eeq
and a similar calculation can be done for the $\mu \to eee$ process.
Contribution from the pseudoscalar exchange also exists, 
but is ignored since it is assumed to be heavy enough.
Using the lower bound on $h_{\mu \tau}$ obtained from $a_\mu$
one can derive conservative upper bounds on 
the flavor changing/conserving couplings such as 
$(h_{e \mu}, h_{\mu \mu}, h_{e\tau}, h_{\tau \tau},h_{ee})$ 
from the experimental limits of the branching ratios of 
$\tau \rightarrow \mu^{-} e^{+} \mu^{-}$,
$~\tau \rightarrow \mu^{-} \mu^{+} \mu^{-}$, 
$~\mu \rightarrow e \gamma$, 
$~\tau \rightarrow \mu \gamma$ and 
$~\tau \rightarrow \mu^{-} e^{+} e^{-}$, respectively.
In addition, we can also predict the upper bounds on
$Br(\tau \to e^{-} \mu^{+} e^{-})$ and $Br(\tau \to e^{-} e^{+} e^{-})$
by combining the upper bounds on $h_{e \tau}, h_{e\mu}$ and $h_{ee}$.
In Table I, we present experimental bounds on the branching ratios
of the corresponding decay processes for $\tau$ and $\mu$ 
and the upper bounds on the flavor changing/conserving couplings.

We note that the above numerical results crucially depends on 
the mass of the scalar.
The bounds on the flavor changing/conserving couplings get increase 
as the mass of the scalar increases.
When we take $m_\phi$ to be 1 TeV, the bound on $h_{\mu \tau}$ is
given by  $0.127 \lesssim h_{\mu \tau} \lesssim 0.693$.
In Table II, we present the upper bounds on the flavor changing/conserving  
couplings and the branching ratio of lepton flavor violating 
three-body decays of the muon and tau lepton
for $m_\phi=1$ TeV.

In conclusion, we have examined the contribution to the anomalous magnetic
moment of the muon from the exchange of the flavor changing scalar.
By using the recent measurement of $(g-2)_\mu$ by the Muon $(g-2)$
Collaboration, we have obtained a bound on $h_{\mu \tau}$ for a given scalar
mass. 
In the analysis, we assumed that the scalar is lighter than the
pseudoscalar to get the positive contribution
and the heavier generations have larger flavor changing couplings.
Useful constraints on the other flavor changing/conserving  
couplings have been obtained
from the lepton flavor violating decays of the muon and tau lepton,
and bounds on the branching ratios of $\tau \rightarrow 3 e$,
$\mu \rightarrow 3 e$ and $\tau \rightarrow e \mu e$ 
have also been predicted.


\begin{table}
\caption{Bounds on the lepton flavor changing/conserving couplings
and branching ratios
  generated from the lepton flavor violating decays of $\tau$ and $\mu$ for 
$m_\phi=100$ GeV.}
\begin{tabular}{ccc} 
 Decay process & Experimental limit & Bound \\ \hline
$\tau \rightarrow e \gamma $ & $2.7 \times 10^{-6}$ & -  \\
$\tau \rightarrow \mu \gamma $ &$ 1.1 \times 10^{-6}$ & 
$h_{\tau \tau}\lesssim 0.113 $\\
$\mu \rightarrow e \gamma $ &$ 1.2 \times 10^{-11}$ & 
$h_{e \tau}\lesssim 6.14\times10^{-5} $\\

$\tau \rightarrow \mu^{-} \mu^{-} \mu^{+}$ &$ 1.9 \times 10^{-6}$ & 
$ h_{\mu \mu}\lesssim 0.0029$ \\
$\tau \rightarrow \mu^{-} \mu^{-} e^{+}$ & $1.5 \times 10^{-6}$ & 
$ h_{e\mu}\lesssim 0.0026$ \\
$\tau \rightarrow e^{-} \mu^{-} e^{+}$ & $1.7 \times 10^{-6}$ & 
$ h_{ee}\lesssim 0.0028$ \\
$\tau \rightarrow e^{-} e^{-} e^{+}$ & $2.9 \times 10^{-6}$ & 
$ Br\lesssim 2.35\times 10^{-12} $ \\
$\tau \rightarrow e^{-} e^{-} \mu^{+}$ & $1.5 \times 10^{-6}$ & 
$ Br \lesssim 2.07 \times 10^{-12} $ \\
$\mu \rightarrow e^{-} e^{-} e^{+}$ & $1.0 \times 10^{-12}$ & 
$ Br \lesssim 2.34 \times 10^{-10}$ 
\\ 
\end{tabular}
\end{table}

\begin{table}
\caption{Bounds on the lepton flavor changing/conserving couplings
and branching ratios
  generated from the lepton flavor violating decays of $\tau$ and $\mu$ for 
$m_\phi=1$ TeV.}
\begin{tabular}{ccc} 
 Decay process & Experimental limit & Bound \\ \hline
$\tau \rightarrow e \gamma $ & $2.7 \times 10^{-6}$ & -  \\
$\tau \rightarrow \mu \gamma $ &$ 1.1 \times 10^{-6}$ & -
\\
$\mu \rightarrow e \gamma $ &$ 1.2 \times 10^{-11}$ & 
$h_{e \tau}\lesssim 0.0026 $\\

$\tau \rightarrow \mu^{-} \mu^{-} \mu^{+}$ &$ 1.9 \times 10^{-6}$ & 
$ h_{\mu \mu}\lesssim 0.122$ \\
$\tau \rightarrow \mu^{-} \mu^{-} e^{+}$ & $1.5 \times 10^{-6}$ & 
$ h_{e\mu}\lesssim 0.108$ \\
$\tau \rightarrow e^{-} \mu^{-} e^{+}$ & $1.7 \times 10^{-6}$ & 
$ h_{ee}\lesssim 0.115$ \\
$\tau \rightarrow e^{-} e^{-} e^{+}$ & $2.9 \times 10^{-6}$ & 
$ Br\lesssim 7.12\times 10^{-10} $ \\
$\tau \rightarrow e^{-} e^{-} \mu^{+}$ & $1.5 \times 10^{-6}$ & 
$ Br \lesssim 6.28 \times 10^{-10} $ \\
$\mu \rightarrow e^{-} e^{-} e^{+}$ & $1.0 \times 10^{-12}$ & 
$ Br \lesssim 7.10 \times 10^{-8}$  
\\ 
\end{tabular}
\end{table}

\begin{thebibliography}{99}

\bibitem{muon}
Muon $(g-2)$ Collaboration, H. N. Brown {\it et al.}, hep-ex/0102017.

\bibitem{amu}
T. Kinoshita and W. J. Marciano, in {\it Quantum Electrodynamics}, edited by
T. Kinoshita (World Scientific, Singapore, 1990), pp. 419-478.

\bibitem{marciano}
A. Czarnecki and W. J. Marciano, hep-ph/0010194.

\bibitem{proposal} 
K.~Lane, hep-ph/0102131;
L.~Everett, {\it et al.}, hep-ph/0102145;
S.~Komine, {\it et al.}, hep-ph/0102204;
J.~L.\ Feng and K.~T.~Matchev, hep-ph/0102146;
E.~A.\ Baltz and P.Gondolo, hep-ph/0102147;
T. Ibrahim, {\it et al.}, hep-ph/0102324;
J.~Ellis, {\it et al.}, hep-ph/0102331;
R.~Arnowitt, {\it et al.}, hep-ph/0102344;
D.~Choudhury {\it et al.}, hep-ph/0102199;
K.~Cheung, hep-ph/0102238;
U.~Mahanta, hep-ph/0102176;
D.~Chakraverty {\it et al.}, hep-ph/0102180;
T.~Huang, {\it et al.}, hep-ph/0102193;
S.~N.\ Gninenko and N.~V.~Krasnikov, hep-ph/0102222;
Z.~Xiong and J.~M.~Yang, hep-ph/0102259;
P.~Das,{\it et al.}, hep-ph/0102242;
T.~W.~Kephart and H.~Pas, hep-ph/0102243;
A.~Dedes and H.~E.~Haber, hep-ph/0102297;
E.~Ma and M.~Raidal, hep-ph/0102255;
Z.~Xing, hep-ph/0102304;
J.~Hisano and K. ~Tobe, hep-ph/0102315;
X.~Calmet, {\it et al.}, hep-ph/0103012;
M.~B. Einhorn and J.~Wudka, hep-ph/0103034;
K.~Choi, {\it et al.}, hep-ph/0103048;
R.~Diaz, {\it et al.}, hep-ph/0103050.

\bibitem{sher1} 
\prd{S.~Nie and M. Sher}{58}{097701}{1998}.

\bibitem{sher2} 
\prd{T.~P.~Cheng and M. Sher}{35}{3484}{1987}.

\bibitem{sher3} 
\prd{M. Sher and Y. Yuan}{44}{1461}{1991}.



\end{thebibliography}
\end{document}